\documentclass[9pt,conference]{IEEEtran}
\usepackage[top=2.1cm, bottom=2.8cm, left=1.65cm, right=1.65cm]{geometry}
\usepackage{cite}
\usepackage{hyperref}
\usepackage{amsmath,amssymb,amsfonts,nccmath}
\usepackage{algpseudocode}
\usepackage{adjustbox}
\usepackage[linesnumbered,ruled,vlined]{algorithm2e}
\usepackage{multirow}
\usepackage[table,xcdraw]{xcolor}
\usepackage{textcomp}
\usepackage{xcolor}
\usepackage{graphicx}
\usepackage[bb=px]{mathalfa} 
\usepackage{tikz}
\usetikzlibrary{positioning, fit, arrows.meta, shapes}
\tikzstyle{line}=[draw]
\usepackage{enumerate}
\usepackage{longtable}
\usepackage{latexsym}
\usepackage{url}
\usepackage[switch]{lineno}
\usepackage[inline]{enumitem}
\usepackage{lipsum}
\usepackage{footnote}
\usepackage{caption}
\usepackage{subcaption}
\usepackage{rotating}
\usepackage{tabto}
\usepackage[utf8x]{inputenc}
\UseRawInputEncoding
\usepackage[linesnumbered,ruled,vlined]{algorithm2e} 

\SetCommentSty{mycommfont}
\usepackage{fancyhdr}

\DeclareRobustCommand*{\IEEEauthorrefmark}[1]{%
  \raisebox{0pt}[0pt][0pt]{\textsuperscript{\footnotesize #1}}%
}

\def\BibTeX{{\rm B\kern-.05em{\sc i\kern-.025em b}\kern-.08em
    T\kern-.1667em\lower.7ex\hbox{E}\kern-.125emX}}
\begin{document}

\title{A YANG-aided Unified Strategy for Black Hole Detection for Backbone Networks}

\author{%
  \IEEEauthorblockN{%
    Elif Ak \IEEEauthorrefmark{1,3},
    Kiymet Kaya \IEEEauthorrefmark{1},
    Eren Ozaltun\IEEEauthorrefmark{1,3}, 
    Sule Gunduz Oguducu\IEEEauthorrefmark{2},
    Berk Canberk \IEEEauthorrefmark{4}
  }%
  \IEEEauthorblockA{\IEEEauthorrefmark{1} Istanbul Technical University, Department of Computer Engineering, Istanbul, Turkey }%
  \IEEEauthorblockA{\IEEEauthorrefmark{2} Istanbul Technical University, Department of Artificial Intelligence and Data Engineering, Istanbul, Turkey }%
  \IEEEauthorblockA{\IEEEauthorrefmark{3} BTS Group, Istanbul, Turkey}
\IEEEauthorblockA{\IEEEauthorrefmark{4} School of Computing, Engineering and Built Environment, Edinburgh Napier University, Edinburgh}
Email: \{akeli, kayak16, ozaltun19, sgunduz\}@itu.edu.tr, B.Canberk@napier.ac.uk}

\maketitle

\thispagestyle{fancy}   
\fancyhead{}                
\lhead{Accepted by 2024 IEEE International Conference on Communications (ICC), \copyright2023 IEEE}
\cfoot{}
\renewcommand{\headrulewidth}{0pt}

\begin{abstract}

Despite the crucial importance of addressing Black Hole failures in Internet backbone networks, effective detection strategies in backbone networks are lacking. This is largely because previous research has been centered on Mobile Ad-hoc Networks (MANETs), which operate under entirely different dynamics, protocols, and topologies, making their findings not directly transferable to backbone networks. Furthermore, detecting Black Hole failures in backbone networks is particularly challenging. It requires a comprehensive range of network data due to the wide variety of conditions that need to be considered, making data collection and analysis far from straightforward. Addressing this gap, our study introduces a novel approach for Black Hole detection in backbone networks using specialized Yet Another Next Generation (YANG) data models with Black Hole-sensitive Metric Matrix (BHMM) analysis. This paper details our method of selecting and analyzing four YANG models relevant to Black Hole detection in ISP networks, focusing on routing protocols and ISP-specific configurations. Our BHMM approach derived from these models demonstrates a 10\% improvement in detection accuracy and a 13\% increase in packet delivery rate, highlighting the efficiency of our approach. Additionally, we evaluate the Machine Learning approach leveraged with BHMM analysis in two different network settings, a commercial ISP network, and a scientific research-only network topology. This evaluation also demonstrates the practical applicability of our method, yielding significantly improved prediction outcomes in both environments.

\end{abstract}


\section{Introduction}
The Internet, an inseparable whole of humans' daily lives, presents a complex landscape that constantly evolves. The backbone networks of Internet Service Providers (ISPs) face various challenges, including fluctuating traffic, network intrusions, and attacks. While collaborative advancements in technology by professionals in industry and academia are continually developed to safeguard against such irregularities and vulnerabilities, certain defects remain undetected and, consequently, unresolved. Among these are the so-called \textit{silent failures} or \textit{Black Holes}. A network Black Hole occurs when a router (or similar network device) unexpectedly and silently discards data packets without notifying the sender. These failures are termed \textit{silent} because they leave no trace, making the troubleshooting process particularly challenging. More importantly, the issue does not disrupt the entire network but affects only the corresponding destination, which is the receiver of dropped packages. This characteristic resembles their namesake in space, the \textit{astronomical black holes}, known for consuming everything around them without a trace.

Black Holes in network systems can arise from several reasons, including hardware malfunctions, setup errors, misconfigurations, or inconsistencies in network protocols. These issues are particularly insidious because they can only be identified by observing the dropped packet flows. Due to the absence of an automatic alert system for Black Holes, they remain undetected unless reported by the affected ISP customers. Without such reports, Black Holes could continue indefinitely in these networks, disrupting communication and data flow without resolution.

The phenomenon of Black Holes is not new to the realm of Internet technology. Despite their long-standing presence, effective strategies to promptly detect them, especially in backbone networks, remain insufficient. Previous research primarily focused on MANETs, which differ significantly from backbone networks regarding network topologies, requirements, and even network protocols. Consequently, findings from MANETs do not readily translate to the backbone networks, restricting the scope of applicable research insights. While there are some initiatives toward Black Hole detection in backbone networks, these are often limited to narrow scenarios, neglecting a broader spectrum of potential network Black Hole incidents \cite{9852788}.

Addressing this gap, our study presents a comprehensive approach to Black Hole detection in backbone networks thanks to specialized data models based on YANG \cite{rfc8345}. YANG data models are increasingly being utilized for network monitoring and autonomous control systems for various case studies, particularly due to their vendor-agnostic approach. However, YANG data models are not specifically designed for black hole detection; instead, they offer a broad perspective for monitoring backbone networks. Therefore, our approach involves a thorough analysis of YANG data models, from which we have selected four that are particularly relevant to Black Hole detection in ISP networks. These models focus on routing protocols and ISP-specific configurations strongly related to Black Hole causes. Throughout the paper, we detail our analysis of these YANG models and demonstrate how the network metrics derived from them significantly enhance prediction performance. This improvement, in turn, leads to an increased packet delivery rate (PDR), showcasing the effectiveness of our approach in addressing this critical network security issue. The contributions of the proposed novel approach are as follows:

\begin{figure*}[ht!]
    \centering
    \includegraphics[width=.75\linewidth]{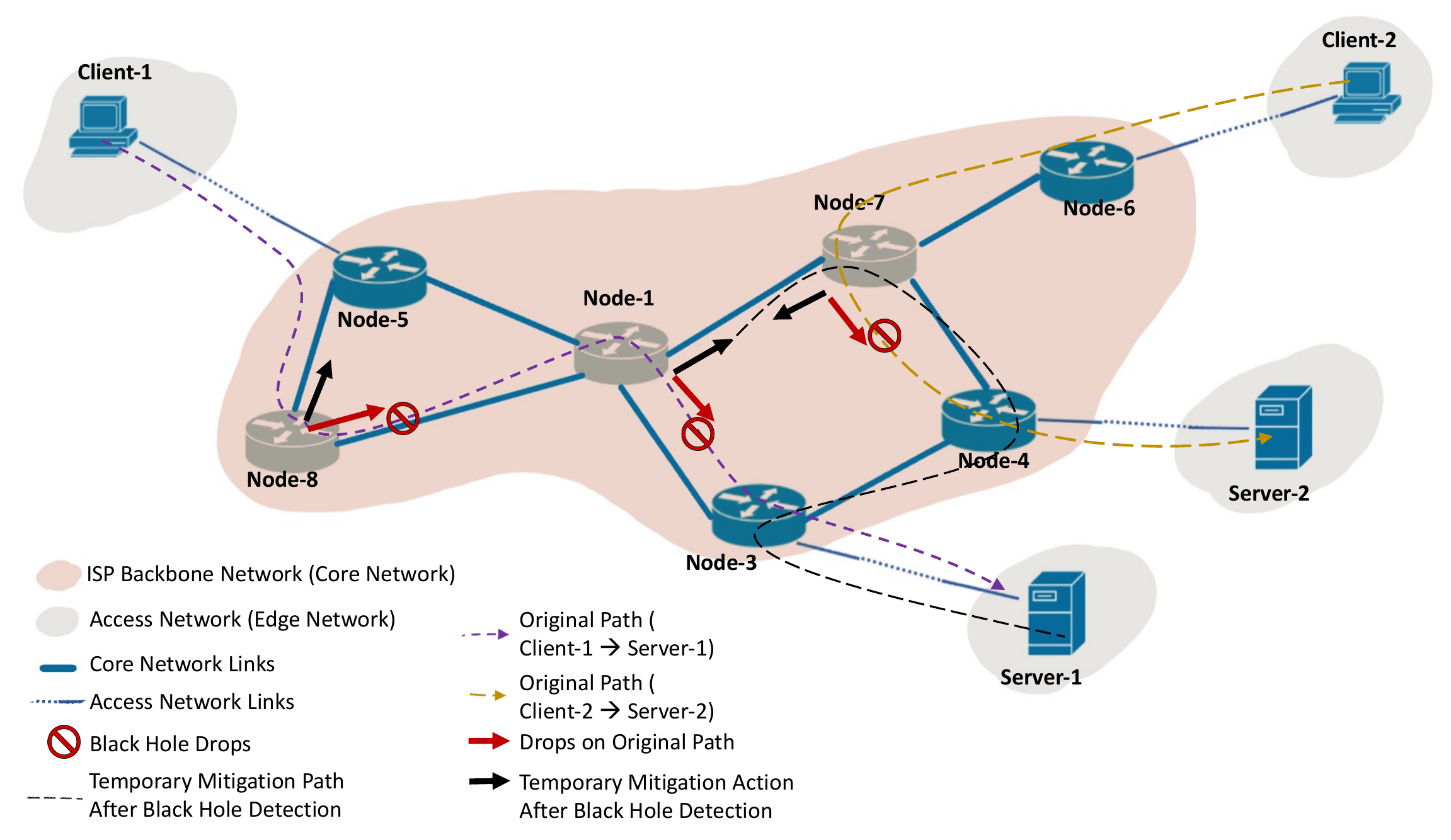}
    \caption{Scientific research-only network topology}
    \label{topology}
\end{figure*}

\begin{itemize}
    \item While YANG data models are not specifically designed for Black Hole detection, our study extensively analyzes and adapts them for this purpose. We carefully select four YANG models that are highly relevant to detecting Black Holes in ISP networks and produce a Black Hole-sensitive Metric Matrix (BHMM).
    \item Through detailed analysis of the chosen four YANG models, we demonstrate how specific network metrics in BHMM can significantly enhance the detection of Black Holes. The pairwise analysis of BHMM results in more accurate detection, showing an average improvement of 10\% in identifying Black Hole failures, with five times less processing time. 
    \item As a result of our proposed approach, we observe an increased PDR with an average 13\% gain. This improvement is a direct consequence of identifying and addressing Black Holes, thus mitigating their impact on network communication and data flow.
    \item Our study fills a critical research gap by focusing on backbone networks, which have different characteristics and requirements compared to MANETs. We provide new insights and solutions tailored to the specific challenges of backbone network Black Hole detection.
    \item The proposed approach is evaluated in two distinct network environments: a commercial ISP network and a specialized network topology exclusive to scientific research. This diversity in application underscores the flexibility and adaptability of the proposed YANG-based approach, enhancing its reliability and practical relevance.  
\end{itemize}

The rest of this paper is organized as follows: Section \ref{sec:rel} reviews current literature relevant to Black Hole detection studies. Section \ref{sec:met} outlines our proposed Black Hole Detection Model through two different network topologies. Section \ref{sec:eval} evaluates our approach using varying sizes of training datasets and provides a detailed analysis of the classification results across different packet flows. Finally, the paper is concluded in Section \ref{sec:conc}, covering future works.

\section{Related Work} \label{sec:rel}

In computer networks, MANETs represent a decentralized structure in which any node can join or leave the network without the need for any permission. In this type of network architecture, there is no client/server structure and is preferred for its ease of use. The flexibility of MANETs, which is seen as an advantage, can also cause the network to be easily exposed to intrusion problems by making it difficult to control and monitor \cite{7387397}. Muneer Bani Yassein et al. proposed a two-stage model for black hole attack estimation in MANETs \cite{yasin2016feature}, which first performs feature selection and then performs classification with J48 tree. In the study of Yaser et al.\cite{khamayseh2019intelligent}, a hybrid routing protocol that optimizes Ad-hoc-on-Demand Distance Vector (AODV) for MANETs with PDR, Dropped Packets Ratio and Average End-to-End Delay features has been proposed for black hole detection. Shweta et al. predicted black hole attacks on MANETs \cite{pandey2020blackhole} using the AODV as the on-demand routing mechanism. They preferred SVM as a prediction method and suggested the SVM-AODV (SAODV) structure that optimizes the routing protocol. For black hole prediction in Wireless Ad hoc Networks (WANETs) \cite{nagalakshmi2021machine} and for black, gray hole, flooding, and scheduling attacks predictions in Wireless Sensor Networks (WSNs) \cite{almomani2016wsn}, ML methods were applied. It has been shown that the prominent classification models can vary according to different evaluation metrics, and it is recommended to choose a metric suitable for the purpose \cite{9525087}. As far as our knowledge, this is the first study that leverages YANG data models for Black Hole detection in backbone networks, considering the features of different routing protocols.

\begin{figure*}[ht!]
    \centering
    \includegraphics[width=\linewidth]{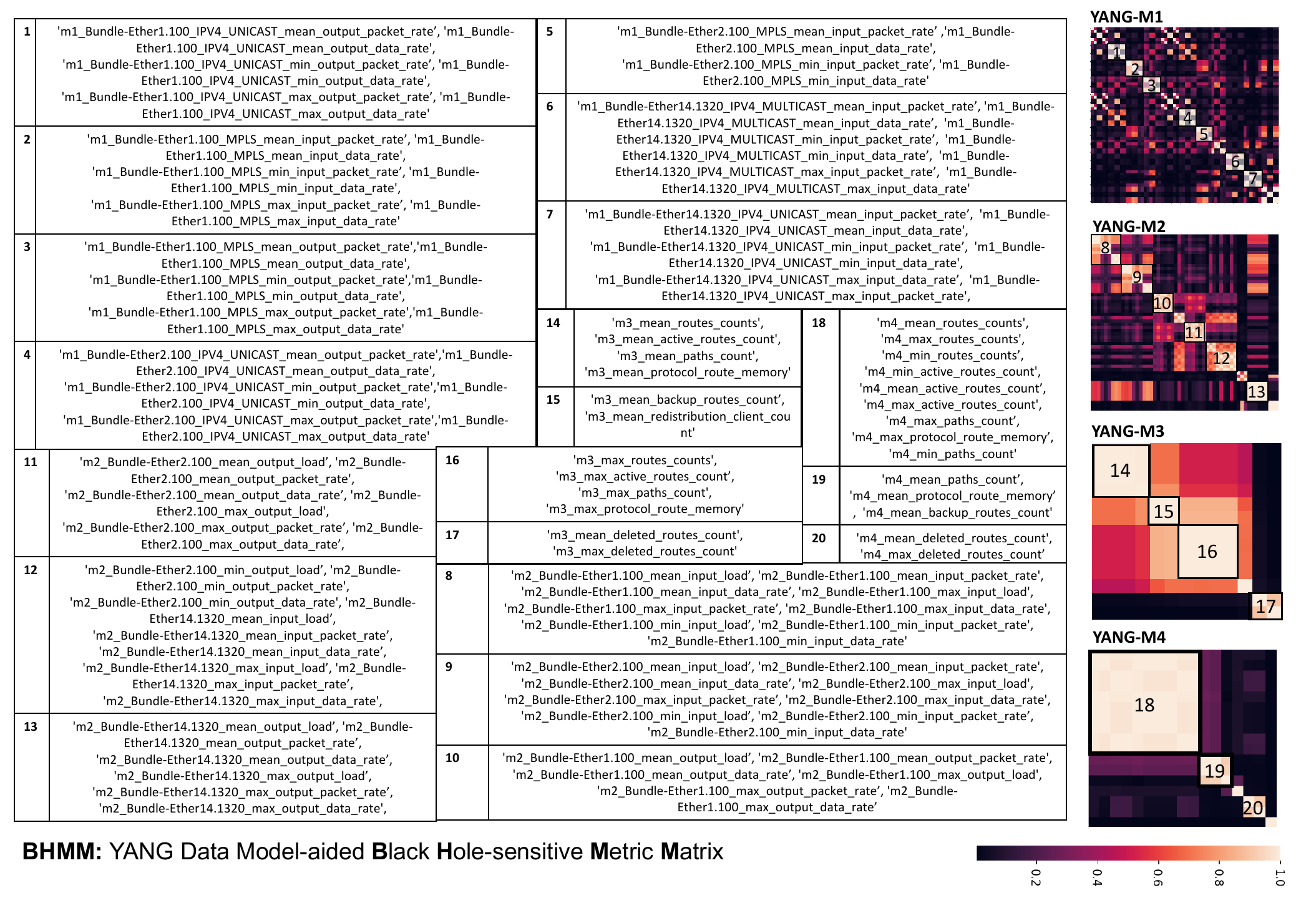}
    \caption{The proposed YANG Data Model-aided Black Hole-sensitive Metric Matrix, called BHMM, for Pairwise Correlation Analysis}
    \label{corr_all}
\end{figure*}

\section{The Proposed Black Hole Detection Model} \label{sec:met}
In this study, we examine two distinct topologies to assess the effectiveness of the proposed YANG models and BHMM analysis when used alongside a well-trained ML model. The first topology under investigation is an actual ISP backbone network, which incorporates various essential protocols, including those for routing. In this real-world ISP environment, YANG models are active within the routers, yet there is no definitive way to ascertain the presence of Black Holes. This scenario aligns with the nature of Black Hole failures, which are inherently silent. We carefully monitored the outcomes generated by unsupervised ML; therefore, we have focused on applying the YANG models and BHMM analysis to this live network setup. The details of the results will be further elaborated in Section \ref{sec:eval}.

To evaluate our approach with the labeled case, where we know the exact time and places Black Hole occurs, we've depicted the topology of our second use case (following the real ISP topology as the first one) with designated Black Hole routes. As shown in Fig. \ref{topology}, this involves data flows from Client-1 to Server-1 and from Client-2 to Server-2. In this research-specific network topology, each router within the ISP backbone network is subjected to a Black Hole occurrence with a probability of 10\%. Specifically, we focus on Node-8, Node-1, and Node-7 for our Black Hole analysis. We chose these nodes because they are centrally located in the ISP network, as opposed to edge network nodes like Node-3, Node-4, Node-5, and Node-6, which have no alternative options to mitigate Black Hole paths.

It is important to note that, in both scenarios, we employed the same YANG models and BHMM analysis to demonstrate their efficacy in enhancing the Machine Learning model's performance and subsequently reducing the PDR. 

\subsection{Black Hole specific Network Packet Flows Capturing}
The YANG, introduced in RFC 7950, is a data modeling language used to model configurational data, operational data, state data, and Remote Procedure Calls (RPCs), generally manipulated by the NETCONF protocol. Since the YANG modeling language is designed to operate on multi-vendor devices, RFC 8345\cite{rfc8345} specifies the YANG data model for network topologies independently from networking companies and their products. 

In the preliminary phase of our study, we identified two YANG data models as particularly relevant to ISP black hole detection objectives. The first model, Cisco-IOS-XR-infra-statsd-oper \cite{yang1}, and the second, Cisco-IOS-XR-ip-rib-ipv4-oper \cite{yang2}, will serve as the foundation for black hole monitoring sensors, constructed using the following YANG paths. We have designated specific sensor groups for the detection of network black holes. These sensor groups, including one network metric, correspond to a feature in ML models.
The used sensor groups are detailed as follows (for simplicity, each sensor group is called with YANG-\textbf{M\#}):
    
\begin{itemize}
    \item \textbf{(YANG-M1)}\textit{Cisco-IOS-XR-infra-statsd-oper:infra-statistics /interfaces/interface/latest/protocols/protocol} models router's Bundle Ethernet interfaces grouped by following protocols: IPV4\_MULTICAST, IPV4\_UNICAST, Multiprotocol Label Switching (MPLS). Among others, this study uses the below sensors from the YANG-M1 model for black hole detection:
    \begin{itemize}
        \item Input Data Rate: It is measured in 1000 bits per second (bps) (or simply 1 kbps).
        \item Input Packet Rate: This refers to the number of packets being received by the interface per second. Unlike data rate, which is concerned with the volume of data, packet rate deals with the number of individual packet units.
        \item Output Data Rate: Like the input data rate, it is measured in kbps. 
        \item Output Packet Rate: This is the number of packets that the interface sends out per second. 
    \end{itemize}
    \item \textbf{(YANG-M2)}\textit{Cisco-IOS-XR-infra-statsd-oper:infra-statistics /interfaces/interface/latest/data-rate} sums up total traffic without protocol details, grouped by Bundle Ethernet interfaces like YANG-M1. We use the same sensors as in YANG-M1, plus:
    \begin{itemize}
        \item Input Load: It measures how much of the interface's total available bandwidth is being consumed by the data it's receiving. This metric is a load as a fraction of 255.
        \item Output Load: It is similar to the input load and with the same unit in a load as a fraction of 255.
    \end{itemize}
    \item \textbf{(YANG-M3)}\textit{Cisco-IOS-XR-ip-rib-ipv4-oper:rib/vrfs/vrf/afs/af /safs/saf/ip-rib-route-table-names/ip-rib-route-table-name/protocol/bgp/as/information} shows the total number of packets related with the Border Gateway Protocol (BGP) protocol without interface breakdown and used sensors are as follows:

    \begin{itemize}
        \item Routes Count: This metric indicates the total number of routes recognized by the BGP from all sources, including active, backup, and deleted.
        \item Active Routes Count: This count refers to the number of routes currently in use or being advertised out to other routers. These are the preferred paths for sending network traffic and indicate the network's current reachability.
        \item Backup Routes Count: Backup routes are not advertised but are kept in reserve to ensure network resilience and continuous data flow.
        \item Deleted Routes Count: Tracking this helps in understanding routing table dynamics and can indicate configuration changes, route optimizations, or responses to network events that might be related to black hole causes.
        \item Paths Count: Unlike route counts, which are about destinations, path count deals with the various paths that packets can take to reach a particular destination, highlighting the network's redundancy and flexibility.
        \item Protocol Route Memory: Monitoring this helps ensure that the router has sufficient resources to operate efficiently, as insufficient memory could lead to route loss or other performance issues.
        \item Redistribution Client Count: Redistribution is used to share routes learned by one routing protocol with another, helping in network integration and route management.
        \item Protocol Clients Count: This metric indicates the number of routing protocol clients connected to the BGP. These clients are other routers or entities that the BGP communicates with to exchange routing information, ensuring network-wide consistency in route knowledge.
    \end{itemize}
    \item \textbf{(YANG-M4)}\textit{Cisco-IOS-XR-ip-rib-ipv4-oper:rib/vrfs/vrf/afs/af /safs/saf/ip-rib-route-table-names/ip-rib-route-table-name/protocol/isis/as/information} is same like YANG-M3 but for Intermediate System - Intermediate System (IS-IS) protocol. We use the same sensors as in YANG-M3 but collected for IS-IS protocol instead of BGP.
\end{itemize}

Moreover, we have monitored the underlying ISP topology in 5-minute periods covering the date range from 01-07-2021 to 30-08-2021 through given YANG models and corresponding network sensors.

\begin{figure*}[ht!]
    \centering
    \includegraphics[width=0.90\linewidth]{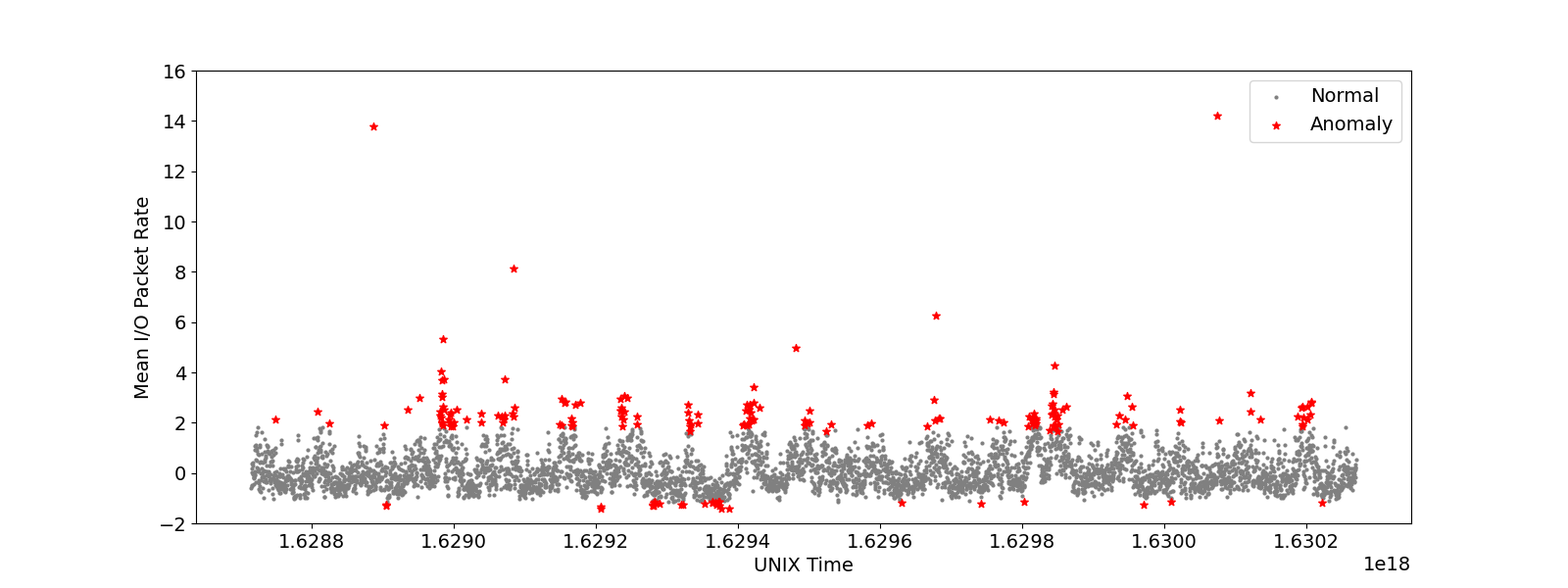}
    \caption{Black Hole detection results for Bundle Ethernet I/O Interface Packet Rate}
    \label{dbscan_io_star}
\end{figure*}

\subsection{The proposed YANG Data Model-aided Black Hole-sensitive Metric Matrix (BHMM) Analysis}
Once YANG paths are monitored within the given duration, the initial dataset created for this study is time series data without any labels. The attribute counts for each YANG model (M1, M2, M3, M4) are 108, 54, 24, and 24, respectively, contributing to a total of 17,280 YANG sensors in the dataset. The feature selection steps are conducted step-by-step as follows:

\subsubsection{Removal of Non-Informative YANG Sensors} We identified and dropped sensors that had only one value across all instances. These sensors were uninformative because they offered no variability or meaningful information for analyzing or predicting black hole occurrences.

\subsubsection{YANG Sensors Sparsity} We addressed the issue of sparsity by identifying and removing features with a high proportion of zero values, which could otherwise skew our machine learning models' performance. At the YANG-M2 sensors, \textit{Bundle-Ether14.1320\_min\_input\_packet\_rate} and \textit{Bundle-Ether14.1320\_min\_output\_packet\_rate} are discarded. 

\subsubsection{Adding Temporal Features} Seasonality is important in black hole flow since the natural reason is unknown and might be affected by anything \cite{10018337}. The black hole flow in the network also varies depending on time. Usually, during the day, there is more data flow than at night, or the data flow will be different on weekdays and weekends. Therefore, we added \textit{minute}, \textit{week of the year}, and \textit{day of the week} features to ensure temporality.

\subsubsection{Adding Black hole Related Features} Input and output packet amounts of black hole routers are inconsistent. Lots of packets reach the router, but the router drops them instead of forwarding them to the destination. Therefore, the amount of output packets is considerably less than the input. Considering this characteristic of the black hole, new features were created by dividing the input by the output on the features related to the input and output, and they were expressed in the study with the \textit{I/O} prefix.

\subsubsection{YANG Paths Correlation Analysis} 
To understand the inter-feature relationships better, we delved deeper into the correlation matrix. We searched for features with a correlation value higher than 0.9 and removed one of the feature pairs from the dataset to reduce calculation time and increase accuracy. In the YANG-M2 model, `Data Rate', `Packet Rate,' and `Load' were highly correlated, as shown in Fig. \ref{corr_all}. We retained `Packet Rate' for its richness in information, removing the others to reduce data noise and enhance model precision. In the YANG-M3 in Fig. \ref{corr_all} and YANG-M4 models in Fig. \ref{corr_all}, we found a similar pattern among `Routes Count,' `Active Routes Count,' `Paths Count,' and `Protocol Route Memory'. Here, we decided to keep `Active Routes Count' and `Protocol Route Memory' because they contained more comprehensive information, contributing to a more nuanced understanding of network behaviors. This decision was consistent for both models, ensuring uniformity in the feature set used for analysis.

In a broader perspective, as seen in Fig. \ref{corr_all}, the point in the matrix with the light colors shows high correlations, which means they are highly possible to give some insight into the ML unsupervised training step. Therefore, only features with fewer correlations, shown as dark colors, and the one representative of the high correlations are kept in the detection model.

\section{Performance Evaluation} \label{sec:eval}

To detect black hole failures, we use DBSCAN, which stands out with its compatibility with unlabelled data and anomaly capture capability among ML models. DBSCAN which is a density-based clustering algorithm \cite{deng2020research} selects high-density groups as clusters and is implemented with two primary hyperparameters, namely $Epsilon$ and $Minimum$  $Points$. 

\begin{figure}[htbp]
\centering
\includegraphics[width=\linewidth]{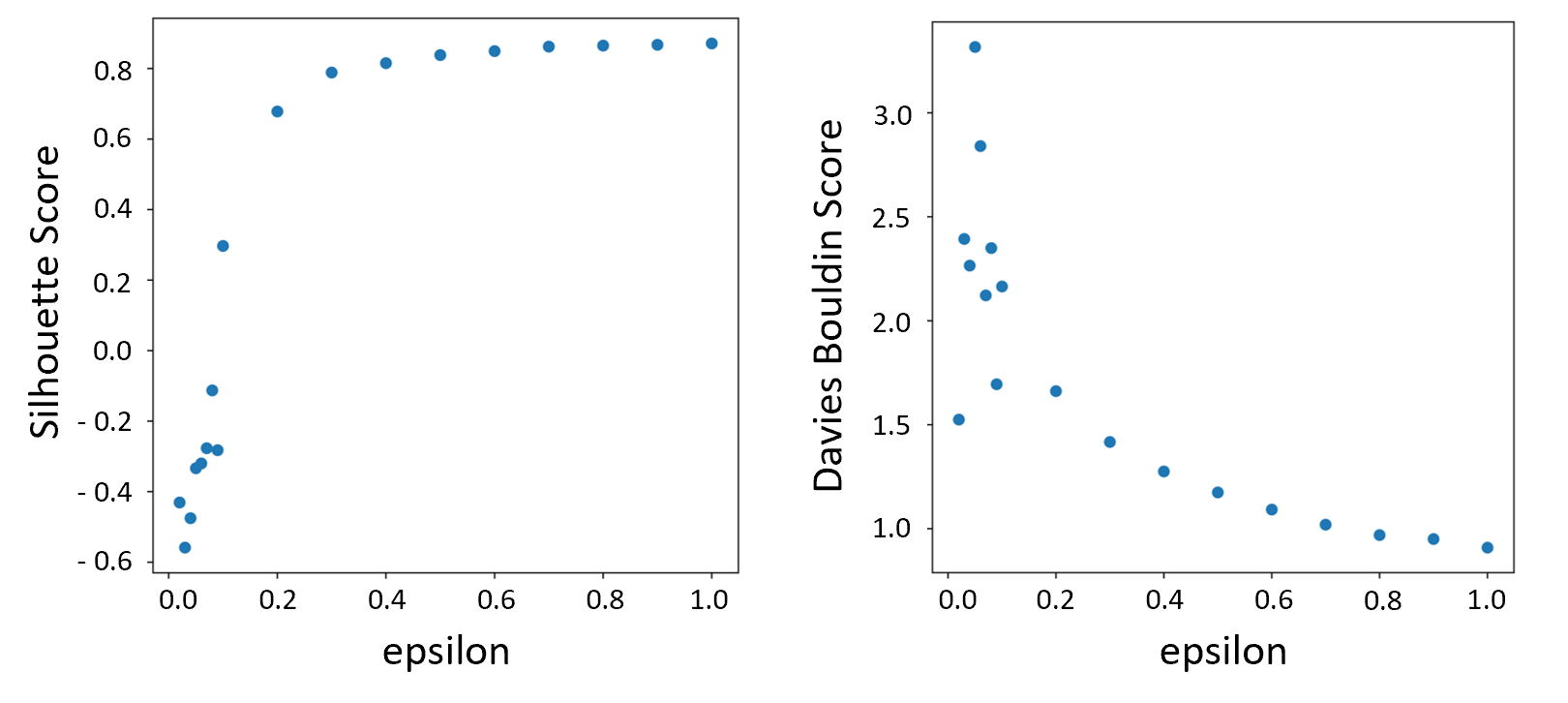}
\caption{DBSCAN Hyperparameter Tuning - Bundle Ethernet I/O Interface Packet Rate}
\label{fig_ss_davies}
\end{figure}
\noindent

These two primary hyperparameters, $eps$, and $minPts$, are tuned according to clustering performance metrics: \textit{Davies Bouldin Score} and \textit{Silhouette Score}. \textit{Davies Bouldin Score} is defined as the average similarity measure of each cluster with its most similar cluster, where similarity is the ratio of within-cluster distances to between-cluster distances. Thus, clusters that are farther apart and less dispersed will result in a better score. The minimum score is zero, with lower values indicating better clustering. \textit{Silhouette Score} is calculated using the mean intra-cluster distance and the mean nearest-cluster distance for each sample and its value ranges from $-1$ to $1$. If the Silhouette Score of clustering is close to $1$, it means that the dataset is well clustered, whereas values close to $-1$ indicate that the samples are assigned to the wrong clusters. Values close to $0$ indicate overlapping clusters. 

\subsection{First Use Case: Commercial ISP Network Data Results}

Since there is a high number of features in the commercial ISP network dataset, for simplicity, we only present DBSCAN results for one feature, \textit{Mean I/O Packet Rate} of \textit{Bundle-Ether1.100}. DBSCAN hyperparameter tuning result for \textit{Mean I/O Packet Rate} of \textit{Bundle-Ether1.100} is shown in Fig. \ref{fig_ss_davies}. Based on the graphs, values higher than $0.8$ are suitable for the hyperparameter $eps$, where the Davies Bouldin Score is low and the Silhouette Score is high. After finding the appropriate $eps$, the same process was repeated for the $minPts$. By continuing these procedures for several iterations, the best values for both hyperparameters of the DBSCAN algorithm were found.

To observe the performance of the DBSCAN, we divided the commercial ISP network dataset into two: the training set consists of $12,096$ instances (70\% of all the data) and the test set consists of $5,184$ instances (30\% of all the data). In addition, $2,592$ samples in training data (15\% of all the data) were used as the validation set. We normalized the data according to standardNorm. We present the well-tuned DBSCAN result for \textit{Mean I/O Packet Rate} of \textit{Bundle-Ether1.100} feature in commercial ISP network data in Fig. \ref{dbscan_io_star}. As seen in Fig. \ref{dbscan_io_star}, the $x$-axis represents the recording time of the samples in the test set of unlabeled black hole data in UNIX time, and the $y$-axis shows the normalized \textit{Bundle Ethernet I/O Interface Packet Rate} feature value of the samples. In Fig. \ref{dbscan_io_star}, the data points in red color represent Black Holes detected via the DBSCAN. In this topology, due to the absence of labels or information about Black Hole occurrences, direct comparison with reality is not possible. However, we can infer potential Black Hole occurrences by comparing anomalies in other network metrics. Limited by space, we've focused on monitoring the Mean I/O Packet Rate and contrasting its anomaly behavior with other metrics. As expected, we found that DBSCAN consistently identifies similar Black Hole events within the same time frames across various network metrics for a given router.

Moreover, we compared the training time of the raw ISP data with the correct feature set of ISP data generated by the BHMM analysis detailed in Section \ref{sec:met}. Computational speed is essential, especially in large networks and real-time applications like the Black Hole detection problem in backbone networks. DBSCAN training in Google Colab, on AMD EPYC 7B12 model CPU with 64 cores and 128 threads, takes 5.059 seconds when 220 features (W/O BHMM) are used for 17280 samples and 1.636 seconds when reduced to 88 features (W/ BHMM).

\subsection{Second Use Case: Specialized Network Topology Results}
We formed training, validation, and test datasets using the same ratios that we applied to our ISP commercial data. The specialized network topology data explicitly includes Black Hole labels. Therefore, we selected 'Accuracy,' 'F1-Macro' (the average F1 score across all classes), and 'Detection Rate' (DR, also known as Recall), which indicates the proportion of actual Black Hole events correctly identified by the model, as our key performance evaluation metrics. Table  \ref{tab:resultsbts} compares the raw results (W/O BHMM) with the outcomes obtained with the correct feature set generation by the BHMM analysis (W/ BHMM) detailed in Section \ref{sec:met} through these evaluation metrics. As it can be seen from the results in Table \ref{tab:resultsbts}, the proposed YANG models aligned with BHMM analysis (green-background color) significantly outperform (W/O BHMM). This outcome is expected, as the BHMM analysis effectively groups network metrics, highlighting key contributors to the ML model. This, in turn, enhances the model's accuracy, F1 Macro, and recall.

\begin{table}[htbp]
\caption{The Effect of BHMM Analysis on Black Hole Detection}
\label{tab:resultsbts}
\centering
\renewcommand{\arraystretch}{1.4}
\resizebox{0.99\linewidth}{!}{
\begin{tabular}{lcccccc}
\hline
 & \multicolumn{3}{c}{\textbf{\begin{tabular}[c]{@{}c@{}}W/ BHMM  Analysis\end{tabular}}} & \multicolumn{3}{c}{\textbf{\begin{tabular}[c]{@{}c@{}}W/O BHMM  Analysis\end{tabular}}} \\ \hline
 & \textbf{Accuracy} & \textbf{F1 Macro} & \textbf{Recall} & \textbf{Accuracy} & \textbf{F1 Macro} & \textbf{Recall} \\ \hline
\textbf{Node 1} & \cellcolor[HTML]{C0E1C1}88.94 & \cellcolor[HTML]{C0E1C1}79.76 & \cellcolor[HTML]{C0E1C1}88.90 & 83.30 & 75.33 & 82.91 \\ \hline
\textbf{Node 7} & \cellcolor[HTML]{C0E1C1}84.56 & \cellcolor[HTML]{C0E1C1}73.17 & \cellcolor[HTML]{C0E1C1}66.71 & 80.12 & 69.66 & 63.71 \\ \hline
\textbf{Node 8} & \cellcolor[HTML]{C0E1C1}89.14 & \cellcolor[HTML]{C0E1C1}79.98 & \cellcolor[HTML]{C0E1C1}89.99 & 83.69 & 74.31 & 80.59 \\ \hline
\end{tabular}
}
\end{table}

\begin{figure}[ht!]
    \centering
    \includegraphics[width=0.85\linewidth]{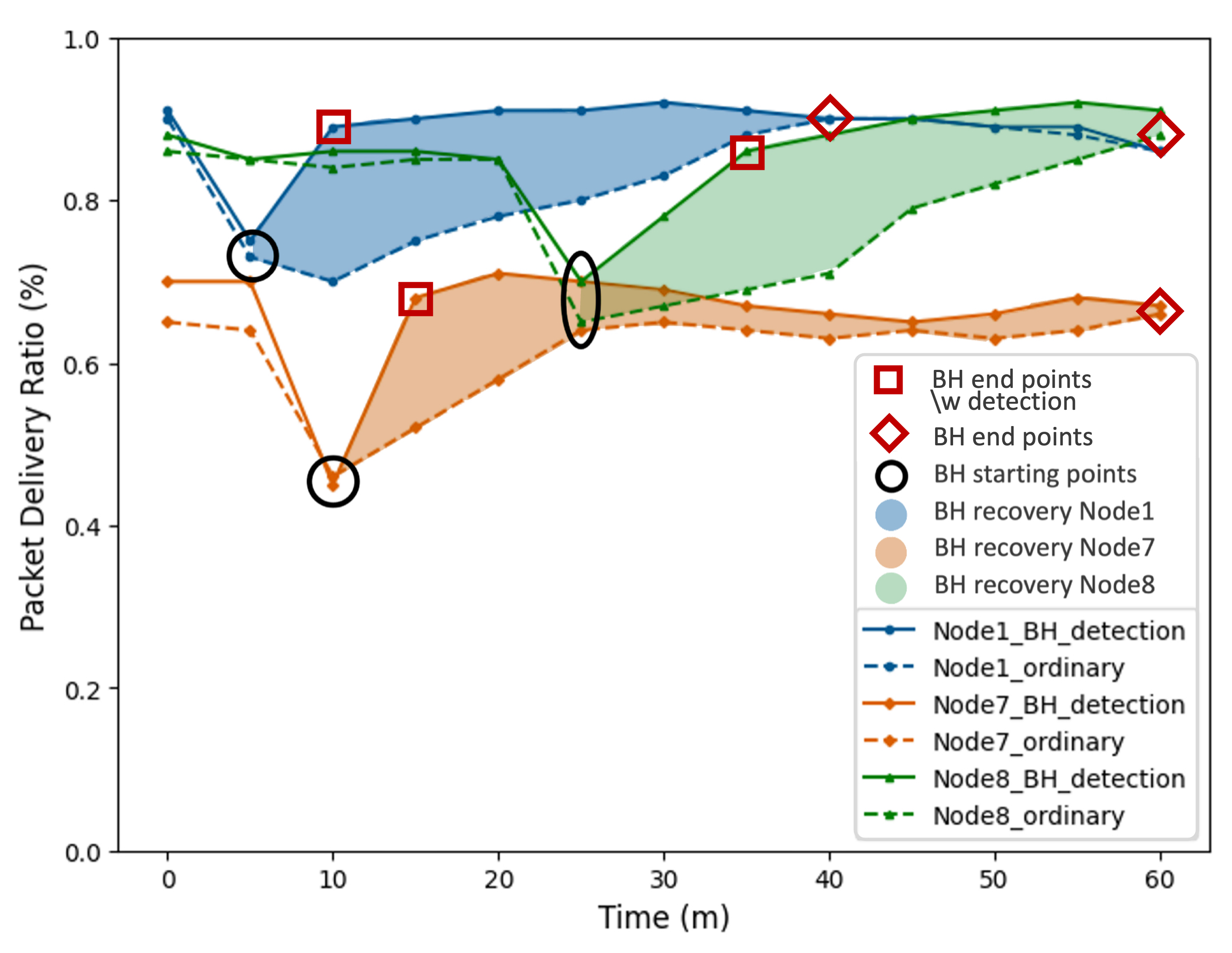}
    \caption{Packet Delivery Ratio (PDR) for Node-1, Node-7 and Node-8 comparing with the proposed detection run or without run }
    \label{pdr}
\end{figure}

Finally, Fig. \ref{pdr} shows how all Black Hole detection approaches and the proposed enhancement might help to increase the PDR. For this analysis, we utilized the scientific-research-only topology as shown in Fig. \ref{topology} and discussed in Section \ref{sec:met}. We focused on Node-1, Node-7, and Node-8, centrally located in the topology, where temporary mitigation strategies are more feasible compared to edge routers. The results are notably promising: the colored areas on the graph represent the mitigation duration for each router. During these intervals, Node-1 and Node-8 experienced Black Holes lasting 15 minutes, while Node-7 had a 5-minute duration. The PDR analysis clearly indicates that applying temporary mitigation after detecting Black Holes significantly improves PDR, with an average increase of 13\%. This improvement is largely attributed to the enhanced detection rate achieved through our proposed use of YANG models and BHMM analysis, ultimately benefiting the PDR for end hosts.

\section{Conclusion and Future Work} \label{sec:conc}
In conclusion, our study represents a significant advancement in detecting Black Holes in backbone networks. By innovatively applying YANG data models, we have developed a Black Hole-sensitive Metric Matrix (BHMM) that not only improves detection accuracy by 10\% but also enhances PDR by 13\%. Our approach effectively addresses the long-standing issue of silent failures in ISP networks, offering a solution that is both practical and adaptable to various network environments. The research contributes to the field by filling a crucial gap in backbone network security, moving away from the limitations of MANET-focused strategies. As a critical area for future work, we plan to investigate strategies for managing False Positive (FP) situations based on a defined algorithmic approach.

\section*{Acknowledgements}
This research is supported by the Scientific and Technological Research Council of Turkey (TUBITAK) 1515 Frontier R\&D Laboratories Support Program for BTS Advanced AI Hub: BTS Autonomous Networks and Data Innovation Lab. project number 5239903, TUBITAK 1501 project number 3220892, and the ITU Scientific Research Projects Fund under grant number MÇAP-2022-43823.

\bibliographystyle{bibliography/IEEEtran}
\bibliography{bibliography/biblio}

\end{document}